# Energy based stochastic model for temperature dependent behavior of ferromagnetic materials

S. Sah[1] and J. Atulasimha[1, a)]

[1]*Department of Mechanical and Nuclear Engineering, Virginia Commonwealth University, Richmond, Virginia 23284, USA*

a) Address correspondence to: *jatulasimha@vcu.edu*

An energy based stochastic model for temperature dependent anhysteretic magnetization curves of ferromagnetic materials is proposed and benchmarked against experimental data. This is based on the calculation of macroscopic magnetic properties by performing an energy weighted average over all possible orientations of the magnetization vector. Most prior approaches that employ this method are unable to independently account for the effect of both inhomogeneity and temperature in performing the averaging necessary to model experimental data. Here we propose a way to account for both effects simultaneously and benchmark the model against experimental data from ~5K to ~300K for two different materials in both annealed (fewer inhomogeneities) and deformed (more inhomogeneities) samples. This demonstrates that the independent accounting for the effect of both inhomogeneity and temperature is necessary to correctly model temperature dependent magnetization behavior.

There have been many magnetization models for ferromagnetic materials such as Preisach model [1-2], Weiss model [3], Stoner-Wolfforth model [4], Brown's analysis of thermal fluctuation in singe domain particles [5], homogenized energy model [6], Jiles-Atherton model [7-8], energy weighted stochastic models [9-13], Globus model [14] other nonlinear constitutive [15] and phase field approaches [16]. While models such as the Preisach model are purely mathematical and do not actually addresses the underlying physics, later models attempt to incorporate specific exchange coupling, shape anisotropy, magnetoelastic anisotropy magnetocrystalline anisotropy and Zeeman energies in describing the magnetization behavior of bulk samples. However, the saturation magnetization, magnetocrystalline anisotropy and average magnetic domain volumes change with a change in temperature. This can present a challenge in modeling temperature dependent magnetization behavior of such materials accurately. In this paper we propose an energy based stochastic approach which can comprehensively model the magnetic behavior of ferromagnetic materials over a range of temperatures by correctly accounting for these temperature effects and benchmark this model against experimental data.

Many prior models approximately employ the following approach [6-14] or some variants thereof to model the magnetization behavior. The total energy density ($E_i$) corresponding to the magnetization orientation along a crystallographic direction ("i") as shown in Fig. 1(a) is evaluated and the probability ($p_i$) of this state being occupied is calculated as:

$$p_i = \frac{e^{-\frac{E_i V}{kT}}}{\sum_i e^{-\frac{E_i V}{kT}}} \qquad (1).$$

Here k demotes the Boltzmann constant, T the temperature and V is the volume as discussed below.

However, there are two challenges in applying such models to modeling ferromagnetic behavior over a wide range of temperature, even if these are well below the Curie temperature ($T_c$): (i) the definition of "V" is nebulous at best for bulk samples and may approximately be assumed to correspond to the average size of a magnetic domain. Even this poses an issue as V may change as domains form, coalesce, etc. during the magnetization process.

(ii) At low temperatures (say when T ~ few Kelvin) this model will only permit the minimum energy states to be occupied that will tend to simulate magnetization curves as shown in Fig. 1 (b), which do not model experimental magnetic behavior at low temperatures correctly.

Some models [9, 10, 13] consider the effect on inhomogeneities (defects, grain boundaries, polycrystalline texture, etc.) on the possibility of occupation of non-minimum energy states. This is an important reason for magnetization cures not looking like Fig. 1 (b) at low temperatures. These models calculate the probability $p_i$ of occupation of state as:

$$p_i = \frac{e^{-\frac{E_i}{\Omega_o}}}{\sum_i e^{-\frac{E_i}{\Omega_o}}} \quad (2).$$

Here, they use an empirical term $\Omega_o$ with no temperature dependence. Hence, both models described by (1) and (2) do not have a framework to model magnetization over a range of temperatures while comprehensively accounting for the effects of magnetic field, magnetocrystalline anisotropy, stress anisotropy, defects, etc. Therefore, we propose to model both effects simultaneously by defining $\Omega$ as follows:

$$\Omega = \Omega_o + \Omega_1 \left(\frac{T}{T_c}\right) \quad (3).$$

Hence, there is an explicit dependence of the occupation of non-minimum states on defects and inhomogeneities (through $\Omega_o$) as well as temperature (through $\Omega_1 (T/T_c)$). It is evident that as $\Omega_o$ increases or $\Omega_1 (T/T_c)$ increases, the high energy states are penalized less (larger denominator) and hence have high probability of being occupied. On the contrary, low $\Omega_o$ (less inhomogeneity) and low $\Omega_1 (T/T_c)$ (low temperatures) would penalize the occupation of high energy states more severely.

The model is implemented by considering the total energy density for the magnetization orienting in a direction ($\alpha_1$, $\alpha_2$, $\alpha_3$) in a cubical anisotropy material due to an applied magnetic field (H) with direction cosine ($\beta_1$, $\beta_2$, $\beta_3$), with respect to crystallographic axes. For this paper, we only consider the magnetocrystalline energy and magnetic energy contributions (we note stress anisotropy and other contributions may be added where appropriate):

$$E = E_{magnetocrystalline} + E_{magnetic}$$
$$= K_1(\alpha_1^2\alpha_2^2 + \alpha_2^2\alpha_3^2 + \alpha_3^2\alpha_1^2) + K_2(\alpha_1^2\alpha_2^2\alpha_3^2)$$
$$-\mu_0 M_s H(\alpha_1\beta_1 + \alpha_2\beta_2 + \alpha_3\beta_3) \quad (4).$$

Where, $K_1$ is the fourth order cubical anisotropy constant, $K_2$ is the second order cubical anisotropy constant, and $M_s$ is the saturation magnetization. The shape anisotropy is not included as the experimental data is corrected for the effects of demagnetization.

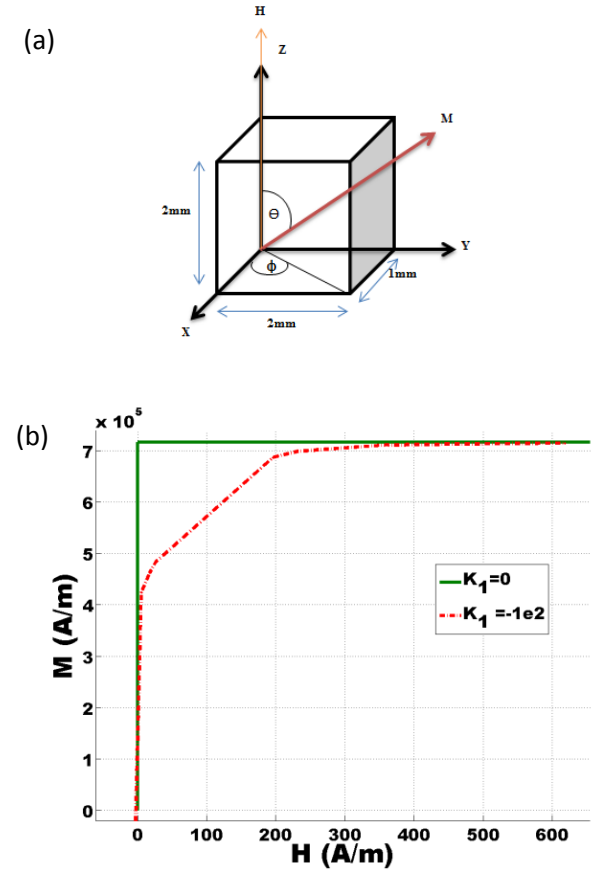

FIG. 1. (a) Schematic of sample (b) M-H curve at 5 K of annealed amumetal sample (at zero and non-zero magneto crystalline anisotropy; with $\Omega_0 = 0$ in both cases).

The total magnetization in the z - direction is then given as [9, 10]:

$$<M_z> = \sum M_s cos\Phi \, P_i$$

$$M_z = \frac{\iint_{(\theta,\phi)} (M_s(T)\sin\theta\cos\phi)(d\theta d\phi)|\sin\theta|)e^{-\frac{E}{\Omega}}}{\iint_{(\theta,\phi)} (d\theta d\phi)|\sin\theta|)e^{-\frac{E}{\Omega}}} \quad (5).$$

Here $M_s$ is the saturation magnetization, $\theta$ is the polar angle and $\phi$ is the azimuthal angle shown in Fig 1 (a).

This new model (equations 3, 4 and 5) only applies to temperatures well below the Curie temperature of the material. The saturation magnetization $M_s$ also depends on temperature. Its dependence on saturation is described as follows [8]:

$$\frac{M_s(T)}{M_s(T_o)} = \left[\frac{T_c - T}{T_c}\right]^a \quad (6).$$

One of the many potential applications of the proposed model is to simulate the behavior of passive ferromagnetic shielding materials that are very important for the proper function of cyomodules of electron beam particle accelerators [17]. The magnetic properties of such materials, subjected to various processing and heat treatment conditions were studied comprehensively over a wide range of temperatures: from cryogenic to room temperature as reported in prior work [18] and the proposed model is therefore benchmarked against this experimental data.

The magnetic characterization in Ref 18 was carried out using Quantum Design VersaLab and a SQUID magnetometer. We obtained experimental data for two Ni-Cr-Fe alloys (amumetal and A4K) at different temperatures (5K to 300K) with different anisotropies in them [18] due to deformation process [19]. For the annealed data, we assume we have very low or vanishing cubic anisotropies in the materials [20]. But the cubical anisotropies are induced in the deformed samples due to permanent deformation that are accounted for with the $K_1$ term.

The following procedure is applied to estimate the model parameters to simulate the behavior of amumetal and A4K bulk samples. The saturation magnetization ($M_s$) was obtained from the experimental data for different temperatures. $K_1(T)$ for the annealed samples was assumed to be zero [19]. $K_2$ for all samples at all temperatures was assumed to be zero. For all annealed samples, $\Omega = \Omega_o + \Omega_1(T/T_c)$ was chosen to give the best fit across all temperatures. For the deformed samples, $\Omega = \Omega_o +$

TABLE I. Modeling parameters $\Omega$ and $K_1$

|  | Undeformed | | Deformed | |
|---|---|---|---|---|
| **Amumetal** | $K_1$ (J/m$^3$) | $\Omega$(J/m$^3$) 418 +(0.21×T) | $K_1$ (J/m$^3$) | $\Omega$ (J/m$^3$) 910 +(0.21×T) |
| 5K | 0 | 419 | -1.59×10$^4$ | 911 |
| 50K | 0 | 429 | -1.51×10$^4$ | 921 |
| 200K | 0 | 460 | -1.09×10$^4$ | 952 |
| 300K | 0 | 481 | -7.06×10$^3$ | 973 |
| **A4K** | | 410 +(0.24×T) | | 880 +(0.24×T) |
| 5K | 0 | 411 | 8.39×10$^3$ | 881 |
| 50K | 0 | 422 | 8.18×10$^3$ | 892 |
| 200K | 0 | 458 | 5.90×10$^3$ | 928 |
| 300K | 0 | 482 | 3.03×10$^3$ | 952 |

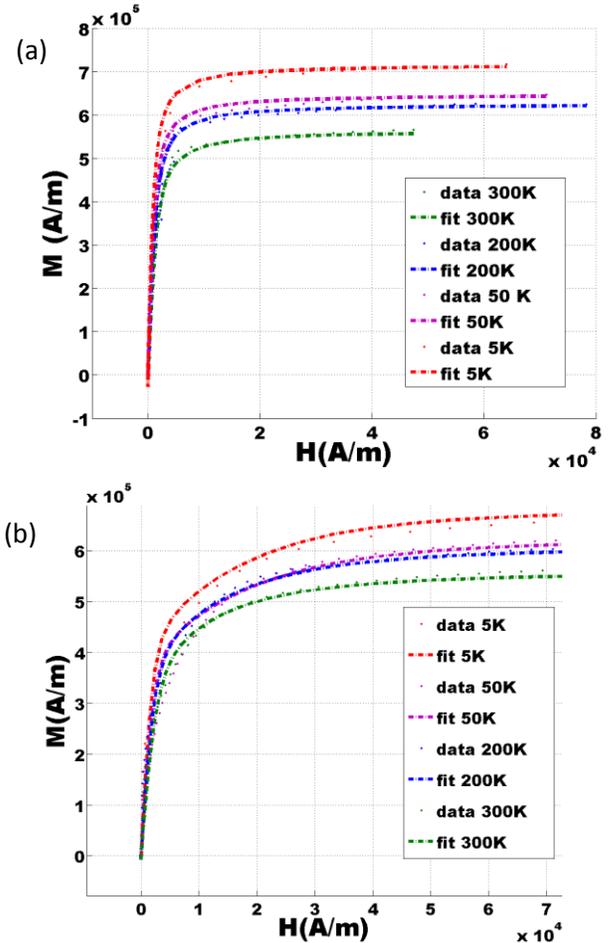

FIG. 2. Comparison between simulated and experimental M-H curves for amumetal samples at different temperatures (a) Annealed amumetal sample with $\Omega = 418 + (0.21×T)$ (b) Deformed amumetal sample with $\Omega = 910 + (0.21×T)$ and empirically chosen $K_1$.

$\Omega_1(T/T_c)$ was chosen to give the initial slope of the curve and $K_1(T)$ was empirically chosen to get the overall best fit. A good correlation was obtained with less than 5% normalized root mean square error in each case as can be seen from Fig. 2 and 3. The model parameters selected for amumetal and A4K are summarized in table I.

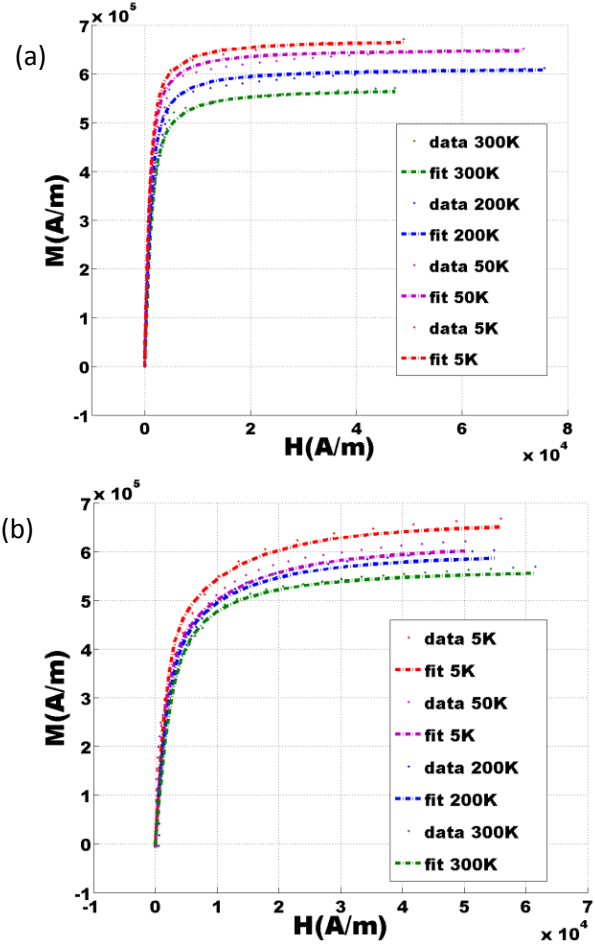

FIG. 3. Comparison between simulated and experimental M-H curves for A4K samples at different temperatures: (a) Annealed A4K sample with $\Omega = 410 + (0.24 \times T)$ (b) Deformed A4K sample with $\Omega = 880 + (0.24 \times T)$ and empirically chosen $K_1$.

Now, we compare the magnetization models using the conventional approach and the new proposed approach for $\Omega$. For the new approach we propose we have:

$$\Omega_i = \Omega_o + \Omega_1\left(\frac{T}{T_c}\right) \quad (7).$$

For the old approach we have:

$$\Omega_{ii} = \frac{KT}{V} = \text{Constant} * (T) \quad (8).$$

For the A4K annealed sample, we compare the results from both approaches and examine the differences. The results for the model using $\Omega_i$ is already presented in Fig. 3. The value of $\Omega_{ii}$ is computed at 5K to find the constant term in equation 8 that would best fit the 5K data. Then the value of $\Omega_{ii}$ is computed at 300K using the constant term and equation 8. The results from both the approaches are plotted against experimental data as shown in Fig. 4 for comparison. While the simulated results from the new approach give us excellent fit with the experimental results at 5 K and 300K, the conventional method's simulated results completely fail to fit the experimental results at 300K. Likewise, if we had tried to fit the 300K data with the conventional approach, the simulated results for the 5K data would have failed to fit the experimental data.

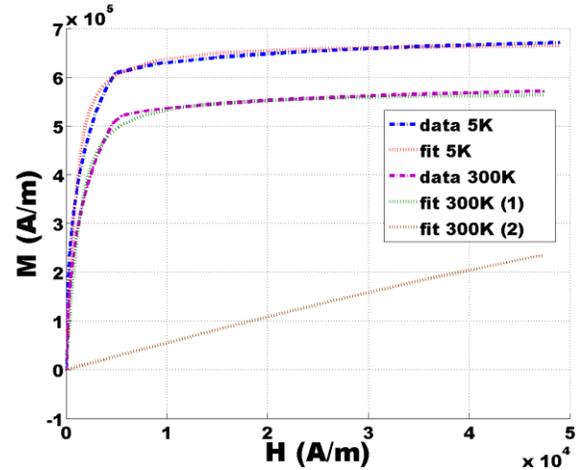

FIG. 4. Comparison of conventional approach and new approach for the simulation of M-H curves of A4K annealed sample at 5K and 300K. This gives the distinct differences between the use of $\Omega_i$ (see fit 300K(1)) and $\Omega_{ii}$ (see fit 300K(2)).

In conclusion, we presented a modified energy based stochastic temperature dependent model that could simulate the magnetization of ferromagnetic materials over a range of temperatures by simultaneously incorporating the effect of inhomogeneity and temperature. As expected, $\Omega_o$ is smaller for annealed samples than deformed samples as the former have less inhomogeneity than the latter, where defects are induced during the deformation process. $\Omega_T=\Omega_1(T/T_c)$ is independent of the processing (i.e. same for annealed or deformed samples) of a given material. This implies that $\Omega_T$ purely models the effect of temperature, independent of the $\Omega_o$ term that only incorporates the effect of inhomogeneity. The $K_1$ (cubic anisotropy) is induced by deformation and its value decreases with increasing temperature as higher temperatures can quench the anisotropy induced by the deformation [21]. In summary, we propose an approach for modeling temperature dependent magnetic behavior of ferromagnetic materials and show that is can simulate the experimental behavior well.

## ACKNOWLEDGEMENTS

We acknowledge collaboration between Virginia Commonwealth University (VCU) and Jefferson Lab (U.S. DOE Contract No. DE-AC05-06OR23177) that partly supports Sanjay Sah.